# RECENT DEVELOPMENTS IN MINIATURIZED OPTICAL SYSTEMS FOR CONTINUOUS FLUORESCENCE DETECTION IN LIQUID FLOWS

**Daniel Mariuta[1,2], Lucien Baldas[1], Jürgen J. Brander[2], Katja Haas-Santo[2], Stéphane Le Calvé[3,4], Stéphane Colin[1], Pascale Magaud[1], Christine Barrot-Lattes[1]**

[1]Institut Clément Ader (ICA), Université de Toulouse, CNRS, INSA, ISAE-SUPAERO, Mines-Albi, UPS, Toulouse France

[2]Karlsruhe Institute of Technology, Institute of Microstructure Technology, Campus Nord, Hermann-von-Helmholtz-Platz 1, 76344 Eggenstein-Leopoldshafen, Germany

[3]University of Strasbourg, Institute of Chemistry and Processes for Energy, Environment and Health (ICPEES), Group of Atmospheric Physical Chemistry, Strasbourg France.

[4]In'Air Solutions, 1 rue Blessig, Strasbourg, France

## SHORT SUMMARY

Miniaturization of continuous fluorescence detection is a challenging task due to the multiple and sensitive parameters intervening in the process. By analyzing fluorescence sensing architectures proposed during the last two years, this work has the goal to identify some trends in the process of fluorescence miniaturization for in-time liquid detection. A lack of postulated strategies regarding the miniaturization process was observed and this review tries to answer partially to this need. The identified integration strategies excel in fulfilling partially the desired functions of a fully autonomous miniaturized detector and further research is needed in order to develop sensing micro-system being capable to step outside of the lab world.

## EXTENDED ABSTRACT

**Introduction**

Today, miniaturized devices follow a straightforward path, and they have come closer and closer to a goal which has been desired for a long time: commercialization of devices with reduced geometrical dimensions and small reagent consumption, being able to perform ultra-high precision analysis to replace the bulky and expensive devices used up to now in many fields such as medicine, biology, chemistry and environmental monitoring. The success in the process of analysis miniaturization will be a historical achievement which will come with a multitude of advantages for the humankind. The costs of analysis which now are expensive from financial, time and logistical point of view would dramatically decrease.

One of the possible analysis methods for these devices is based on fluorescence detection. There are some recent attempts to develop on-chip detection systems, whose contributions and successes maintain the hope that in the near future these devices will cross the border of commercialization. A possible reason for not having yet merchandisable on-chip fluorescence sensing platforms is the lack of postulated modular integration strategies of fluorescence measurement chain in order to further be mastered and improved iteratively.  Within the present work, the novel fluorescence sensing platforms from the last two years have been analyzed for identifying some current on-chip integration strategies of the fluorescence measurement components used for developing ultra-portable, compact and very sensitive detection devices. An ideal integration strategy has to provide  automatic, low cost, multiple analysis, fast response time and compact on-chip sensing platform.

**Literature review**

**1. Printed board circuits (PBC) used as mechanical and electrical support for light emitting sources and/or photon detectors** represents a strategy permitting a high degree of modularity and a parallel layering distribution of the detector device functions. This strategy was used for fluorescence sensing platforms proposed by Babikian et *al.* [1] and Shin et *al.*, [2]. Moreover, both proposed heterogeneous integration strategies/architectures have the potential to be broadly applicable for multiple analysis. The on-chip fluorescence detection device proposed by Babikian et *al.* [1] integrates optical, electronic, electro-mechanical, and fluidic components on the same chip (figure 1, left). Four well established technologies (printed circuit board, microelectronics, isotachophoresis and digital signal processing) were integrated in order to provide a device with the form factor of a credit card. This device avoids using bulky high-quality components and filters, being classified as a lensless fluorescence detection system. The detection sensitivity reached is 10 nM for measuring the concentration of a fluorescein within a microchannel of 500μm×100μm. In the cited work it was demonstrated that PCB can support heterogeneous and 3-D integration within a microfluidic chip which is also important from a manufacturing point of view. Shin et *al.* [2] developed a portable fluorescent sensor platform which employs multiple excitation sources (figure 1, right). The light sources, having different wavelengths, are turned on or off sequentially to stimulate different fluorescent molecules developing in this way a portable multi-analytes fluorescent system for multiple target detection. The light emitting diodes are controlled by an integrated electronic circuit and a microcontroller on board and detection is ensured by a highly sensitive silicon photomultiplier. The disposability of the microfluidic part is a very important feature of this device.

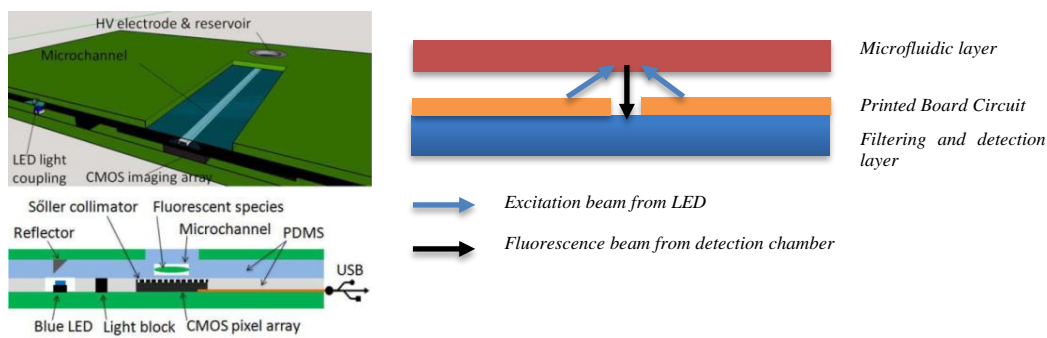

**Figure 1: Left -** Cross-sectional diagram of the microfluidic PCB system, reprinted from [1]. **Right -** Parallel layering integration strategy proposed in [2]

**2.** Ngernsutivorakul et *al.* [3] proposed a **strategy which integrates lenses, filtering networks and optical fibers into an PDMS probe** (figure 2). This strategy might enable a new route which coupled with the progress registered in the field of microfabrication could replace the complex and critical alignment of bulky and expensive optical components, which were drastically limiting the miniaturization process. Polydimethylsiloxane (PDMS) was used to create 0.5 mm thick and 1 mm wide compact probes into which pre-aligned mirrors, lenses and two fiber optic guides are incorporated. The incorporation of the filters at the tip of the probe would diminish the background noise and improve the sensitivity. This probe is connected with the chip in order to facilitate the excitation in the detection chamber and collection of the fluorescence as well. Further improvements are possible according to the authors by using more compact light sources and detectors. The efficiency was tested for different capillary dimensions and for different applications and the limit of detection was found being comparable with the other micro devices using integrated optics and optical fibers (100 nM).

**3. Organic layers used as light sources and photon detectors** is not a new strategy in the miniaturization process of fluorescence detection, but some recent works (Shu et *al.* [4], Jahns et *al.* [5]) reported considerable achievements. Shu et *al.* [4] introduced and tested for the first time a fluorescence light detector combining both an organic electrochemical cell (OLEC) and an organic photodiode (OPD)

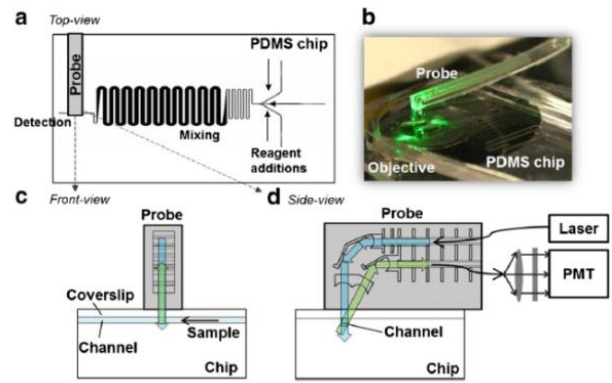

**Figure 2: a.** Top view of the probe positioned on a PDMS chip, **b.** Image on the probe on the PDMS chip, **c.** Front view of the probe on a chip **d.** Side-view of the probe on a chip and the experimental set-up for fluorescence detection. Reprinted from [3]

within one microchip (figure 3). Organic electronics introduce some unbeatable advantages such as direct on-chip integration, easy emission and detection wavelength tuning and compatibility with flexible substrates. The proposed sensor is able to measure fluorescein concentrations lower than 1 µM. Furthermore, fully solution processing and vacuum free fabrication procedures used for fabrication of the OLEC and OPD, easy tuning of the excitation light color enable another promising integration strategy for low-cost, disposable, point of care applications [4].

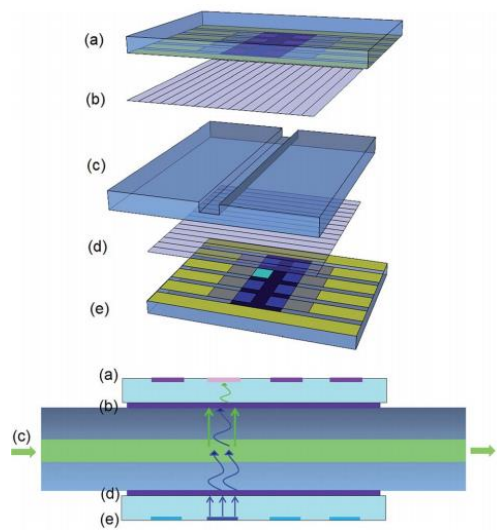

**Figure 3:** a. fully solution processed OPD; b. first linear polarization filter oriented at 0 degree; c. glass microfluidic chip; d. second linear polarization filter oriented at 90 degree; e. fully solution processed blue OLEC. Reprinted from [4].

**4.** Guduru *et al*. [6] proposed an on-chip configuration which integrates in the same glass substrate the microfluidic network, the excitation source, the filtering and the light collection elements (figure 4). The strategy proposed by the authors involves the **implementation of one dimensional photonic crystal layer working similarly as a Bragg mirror for a selective selection of the wavelengths and the fluorescence collection is performed by a Binary Fresnel Lens (BFL) which is placed on top of the photonic crystal.** For the proposed configuration, the photonic crystal, which has the role of a filter, was fabricated on top of the layer containing the microchannel network. The measurement efficiency was not very high due to defects in the silica film, non-total transparency of the ablated regions and irregularities in the Fresnel zones. The authors mentioned that the optical quality of the photonic crystal could be improved using thermal treatment after the deposition process. Higher manufacturing precision would increase the device efficiency.

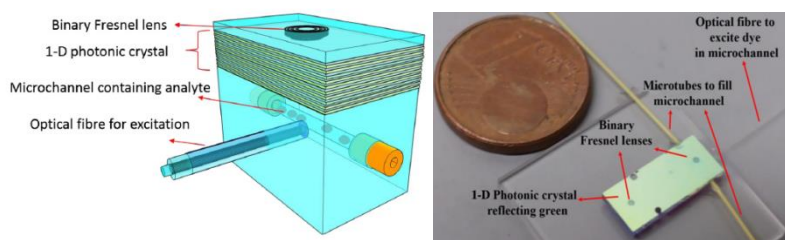

**Figure 4:** Microfluidic chip with the Fresnel lens, composed of a fluidic channel, an excitation fiber perpendicular to it and a filtering and collection part on the top surface**.** Reprinted from [6]

**Conclusion**

During the last years, a considerable effort has been conducted in order to propose viable solutions for multiple target, low cost and sensitive on-chip fluorescence detection. It has been found a lack of reviews in the literature to summarize and point out the new achievements regarding the light emitting sources, waveguides, focusing optics, miniaturized photon detectors and strategies to couple all these into a cheap, easy to use, compact micro-detection system.

Printed board circuits allow development of a modular architecture and the potential for a very compact, reproducible, easy to use multi target analyzers. Further research should be focused on developing methods to efficiently couple the excitation/emission light transfer in order to maximize the signal to noise ratio and avoid direct illumination of the photon detector. The strategy proposing the integration of the filtering, lenses and optical fiber into a micro fabricated PDMS probes has the advantage of drastically miniaturizing the optical elements of the measurement process, but is not suitable for multiple target analysis. Organic layers used as light sources and photon detectors is the most promising integration strategy from the point of view of automatization, compactness, low cost fabrication and ease of use. Long term functionality of the organic layers has to be further improved to maximize the potential for a real autonomous and merchandisable sensing platform. The forth strategy identified within this review, proposes a layering filtration method based on a 1-D photonic crystal and an efficiently fluorescence collection using binary Fresnel lenses, all integrated within two thin layers on a chip. This strategy could be improved in order to increase the ease of use. For example, avoiding the use of optical fiber for excitation light transfer is one of the possibilities, considering the complexity of light alignment issues in this case.

**Acknowledgements**

This project has received funding from the European Union's Framework Programme for Research and Innovation Horizon 2020 (2014-2020) under the Marie Sklodowska-Curie Grant Agreement No. 643095.